\documentstyle[12pt]{article}

\begin{document}

\begin{titlepage}

\title{\large\bf  Collider Signatures of Neutrino  Masses and Mixing 
from R-parity Violation}

\author{ {\normalsize 
           Eung Jin Chun$^a$, Dong-Won Jung$^b$, 
           Sin Kyu Kang$^{c,d}$, Jong Dae Park$^b$}  \\
{\normalsize\it $^a$Korea Institute for Advanced Study} \\
    {\normalsize\it P.O.Box 201, Cheongryangri, Seoul 130-650, Korea}\\
{\normalsize\it $^b$Department of Physics, Seoul National University,
     Seoul 151-747, Korea}\\
{\normalsize\it $^c$Institute for Basic Science, Korea University, 
                  Seoul 136-701, Korea}\\
{\normalsize\it $^d$Graduate School of  Science, Hiroshima University, 
                 Higashi-Hiroshima 739-8526, Japan}
}

\date{}
\maketitle

\abstract{R-parity violation in the supersymmetric standard model
can be the source of neutrino masses and mixing. We analyze the 
neutrino mass matrix coming from either bilinear or trilinear
R-parity violation and its collider signatures, assuming 
that the atmospheric and solar neutrino data are explained by 
three active neutrino oscillations.
Taking the gauge mediated supersymmetry breaking mechanism,
we show that the lightest neutralino decays well inside the detector and the 
model could be tested by observing its branching ratios 
in the future colliders.  In the bilinear model where only the 
small solar neutrino mixing angle can be accommodated, 
the relation, $10^3$ BR($\nu e^\pm \tau^\mp$) 
$\sim$ BR($\nu \mu^\pm \tau^\mp$) $\approx$ 
BR($\nu \tau^\pm \tau^\mp$), serves as a robust test of the model.
The large mixing angle solution can be realized in the trilinear model  
which predicts BR($\nu e^\pm \tau^\mp$) $\sim$ BR($\nu \mu^\pm \tau^\mp$) 
$\sim$ BR($\nu \tau^\pm \tau^\mp$).  In either case, 
the relation, BR($e jj$) $\ll$ BR($\mu jj$) $\sim$ BR($\tau jj$),
should hold to be consistent with the atmospheric  neutrino
and CHOOZ experiments.
}

\end{titlepage}


\section{Introduction}

In the supersymmetric standard model, the gauge invariance and 
renormalizablity allow lepton and baryon number violation and thus it 
may cause too a fast proton decay.   Such a problem is usually avoided by
introducing a discrete symmetry.  Among various possibilities,
the $Z_2$ R-parity and $Z_3$ B-parity have been advocated 
as they can be remnants of gauge symmetries in string theory \cite{ibro}.
Imposing R-parity has been more popular because of its simplicity and the
possibility of having a natural dark matter candidate.
The second option of allowing lepton number violation is also of
a great interest since it can generate neutrino masses and mixing 
\cite{hasu} in an economical way to explain the current neutrino data.   
There is a huge (but incomplete) list of literature investigating 
neutrino properties in this  framework \cite{oldies}.

R-parity violation may lead to a distinctive collider signature that 
the usual lightest supersymmetric particle (LSP), which is typically
a neutralino, produces clean lepton (or baryon) number violating 
signals through its decay \cite{sig91}.  In a model of neutrino masses 
and mixing with R-parity violation,  one can have more specific
predictions for various branching ratios of the LSP decay,
as  the structure of lepton flavor violating couplings is dictated 
by the pattern of neutrino mixing determined from neutrino experiments
\cite{mrv,jaja,valle2}.   This provides a unique opportunity to test
the model in the future collider experiments.
A necessary condition is of course that the LSP has a short lifetime 
to produce a bunch of decay signals inside the detector.
In the models we will consider, the total LSP decay rate is proportional 
to the (heaviest) neutrino mass and thus the measurement of the
LSP decay length could also be useful to test the model.

\medskip

It is the purpose of this paper to examine the correspondence between
neutrino oscillation parameters and collider signatures 
charactering specific models of
neutrino masses and mixing from R-parity violation.
For this, we will consider the bilinear and trilinear models 
to see whether they can accommodate the 
atmospheric \cite{sk-atm} and solar neutrino oscillations \cite{sol-exp}
and the constraint coming from the CHOOZ experiment \cite{chooz}, 
simultaneously.
One of our basic assumptions is the universality of soft supersymmetry
breaking terms at a high scale, which is usually imposed to avoid 
flavor problems in the supersymmetric standard model.  This implies
that  the lepton flavor violation occurs only in the superpotential 
with bilinear and/or trilinear  R-parity violating terms and 
the supersymmetry breaking mechanism is flavor-blind.  Then, 
the tree-level neutrino mass is generated by the renormalization 
group  evolution  which breaks universality between 
the slepton and Higgs soft terms at the weak scale.   
As a specific scheme, we will consider 
the mechanism of gauge mediated supersymmetry breaking 
which solves the supersymmetric flavor problem in a natural way
\cite{gmsb}.  A comprehensive analysis of neutrino masses and 
mixing in this context has been performed 
in Ref.~\cite{CK}.\footnote{For a recent detailed analysis,
see Ref.~\cite{KK}.}

Under such an assumption, the bilinear model can only realize the small
mixing angle of solar neutrino oscillations while the 
trilinear model can accommodate the large mixing angle as well.
In both cases, we will investigate whether the LSP decay length is short
enough and what are the predictions for LSP decay signals which could
test the model in the future collider experiments.
Here, another assumption we make is that the LSP is a neutralino.
Let us remark that a similar analysis has been made 
in Ref.~\cite{valle2} considering supergravity models with
generic bilinear R-parity violating terms. 

\medskip

This paper is organized as follows.  In Sec.~2, we calculate the
``effective'' trilinear R-parity violating couplings,
rotating away the mixing mass terms between the ordinary 
particles and superparticles which arise as a consequence of
bilinear R-parity violation.  Those couplings are relevant for
the LSP decay.  In Sec.~3, we examine the neutrino mass matrix
which is generated through  renormalization group evolution
and various (finite) one-loop diagrams.  From this, we will make
a qualitative analysis to examine the sizes of various R-parity 
violating couplings which are required to explain the current neutrino 
oscillation data.  In Sec.~4, we will provide  a numerical analysis
to determine R-parity conserving and violating input parameters 
with which the atmospheric and solar neutrino masses and mixing are 
realized, in the context of gauge mediated supersymmetry breaking models.
Calculating the corresponding LSP decay rate and branching ratios 
of various modes,  we will find how the model can be tested in the
collider experiments.  Finally, we will conclude in Sec.~5.

\section{Effective R-parity violating vertices from bilinear terms}

Allowing lepton number violation in the supersymmetric standard model,
the superpotential is composed of the  R-parity conserving $W_0$ and 
violating $W_1$ part;
\begin{eqnarray} \label{supo}
 W_0 &= & \mu H_1 H_2 + h^e_i L_i H_1 E^c_i + h^d_i Q_i H_1 D^c_i 
         + h^u_i Q_i H_1 U^c_i  \nonumber\\
 W_1 &=& \epsilon_i \mu L_i H_2 +  {1\over2}\lambda_{ijk} L_i L_j E^c_k
      +    \lambda^\prime_{ijk} L_i Q_j D^c_k \,.
\end{eqnarray}
Among soft supersymmetry breaking terms, let us write R-parity 
violating bilinear terms;
\begin{equation}
 V_{soft} = B\mu H_1 H_2 + B_i \epsilon_i \mu  L_i H_2  
            +  m^2_{L_iH_1} L_i H_1^\dagger + h.c. \,.
\end{equation}
It is clear that the electroweak symmetry breaking gives rise 
to nonzero vacuum expectation values of sneutrino fields, $\tilde{\nu}_i$, 
as follows \cite{hasu};
\begin{equation}
 a_i \equiv {\langle \tilde{\nu}_i \rangle \over 
   \langle H_1 \rangle} =  
     - { \bar{m}^2_{L_iH_1} + B_i \epsilon_i \mu t_\beta 
             \over m^2_{\tilde{\nu}_i} }
\end{equation}
where $\bar{m}^2_{L_iH_1}= m^2_{L_iH_1}+ \epsilon_i \mu^2$, 
$t_\beta=\tan\beta = \langle H_2 \rangle/ \langle H_1 \rangle$ and 
$m^2_{\tilde{\nu}_i}= m^2_{L_i}+ M_Z^2 c_{2\beta}/2$.
In general, there are three types of independent R-parity violating 
bilinear parameters such as $ \epsilon_i$,  $a_i$ and $B_i/B$,
which give rise to the mixing between the ordinary particles and 
superparticles.
That is, neutrinos and neutralinos, charged leptons and charginos, 
neutral Higgs bosons and sneutrinos, as well as, charged Higgs 
bosons  and charged sleptons have mixing mass terms which are 
determined by the above R-parity violating parameters.
The mixing between neutrinos and neutralinos particularly
serves as the origin of the tree-level neutrino masses 
which will be discussed later.
Note that the above quantities have to be very small 
to account for tiny neutrino masses.
While the effect of such small parameters on  
the particle and sparticle mass spectra (apart from the neutrino sector)  
are negligible, they induce small but important R-parity violating vertices 
between the particles and sparticles, which make the LSP destabilized
and generate one-loop neutrino masses.
The derivation of the induced R-parity violating couplings
has been performed in many previous works.
The usual approach is to take full diagonalizations of enlarged
sparticle--particles mass matrices with R-parity violating parts
so that the vertices in terms of the mass eigenstates are obtained 
directly.

In this work, we take an alternative but equivalent approach 
which is useful when R-parity violating parameters are small.  
It is to rotate away only the small R-parity violating
(off-diagonal) blocks of the particle--sparticle mass matrices, leaving
untouched R-parity conserving particle or sparticle masses at 
the diagonal blocks.   In this way, we can draw the induced
(or ``effective'') R-parity violating vertices in terms of the
electroweak/flavor eigenstate basis.  A merit of this method  is that
one can clearly see the vertex structure of the induced R-parity 
violating couplings along with the usual trilinear vertices in $W_1$ of 
Eq.~(\ref{supo}) added to the usual R-parity conserving
Lagrangian.    This is nothing but the usual see-saw 
diagonalization, which we summarize as follows.
Let us take sparticle--particle mass matrix given by
$$ \pmatrix{ M & \Delta \cr \Delta^\dagger & M' } $$
with $\Delta \ll M, M'$. Then, the approximate diagonalization 
(valid up to the second order of R-parity violating parameters $\sim
\Delta/M$ or $\sim \Delta/M'$)
can be done with the help of the rotation matrix  given by
$$ \pmatrix{ 1-{1\over2} \Theta \Theta^\dagger & -\Theta \cr 
              \Theta^\dagger & 1 -{1\over2} \Theta^\dagger \Theta } $$
where $\Theta$ can be found by solving the relation,
$\Delta=M \Theta-\Theta M'$, in the leading order of $\Delta$.  
The upper and lower diagonal blocks are then shifted as 
$M \to M + (\Theta \Delta^\dagger+ \Delta \Theta^\dagger)/2$ and 
$M' \to M' - (\Theta^\dagger \Delta+ \Delta^\dagger \Theta)/2$.
Note that  the neutrino-neutralino mass matrix has vanishing sub-matrix 
for the neutrinos, $M\equiv0$, and the above change in $M$ 
is just the see-saw generation of small neutrino masses.
For the other particles/sparticles, such changes can be safely
neglected.  After  performing such a rotation, we get the 
``effective'' R-parity violating vertices in the 
electroweak/flavor basis. Then, it is quite straightforward to find the 
corresponding couplings in the mass basis following the usual 
diagonalization of the familiar (R-parity conserving) particle/sparticle 
mass matrices.

In this paper, 
we do not repeat to write the mixing mass terms between sparticles
and particles.  Instead,  we will present the rotation matrices $\Theta$ 
in terms of the following three bilinear R-parity violating variables;
$$ \epsilon_i\; (\mbox{or } a_i)\;, 
    \quad \xi_i\equiv a_i - \epsilon_i\;, \quad
    \eta_i \equiv a_i - B_i/B \;. $$
In generic supersymmetry breaking models with non-universality, 
the above three types of parameters are independent.  
But, in the restrictive models imposing the  universality condition
at the mediation scale of supersymmetry breaking,
nonzero values of  $\xi_i$ and $\eta_i$ arise as a consequence of 
renormalization group evolution and thus only two types of parameters
are independent.  In this paper, we usually take 
$\epsilon_i$ and $\xi_i$ as independent ones.

\medskip

\underline{Neutrino-neutralino diagonalization}

Rotating away the neutrino-neutralino  mixing mass terms 
(by $\theta^N$) can be
made by the following redefinition of  neutrinos and neutralinos:
\begin{equation}
 \pmatrix{ \nu_i \cr \chi^0_j } \longrightarrow
 \pmatrix{ \nu_i- \theta^N_{ik} \chi^0_k  \cr
           \chi^0_j + \theta^N_{lj} \nu_l }
\end{equation}
where $(\nu_i)$ and $(\chi^0_j)$ represent three neutrinos
$(\nu_e, \nu_\mu, \nu_\tau)$ and four neutralinos
$(\tilde{B}, \tilde{W}_3, \tilde{H}^0_1, \tilde{H}^0_2)$ 
in the flavor basis, respectively.  The rotation elements
$\theta^N_{ij}$ are given by 
\begin{eqnarray}
 \theta^N_{ij} &=& \xi_i c^N_j c_\beta - \epsilon_i \delta_{j3} 
 \quad\mbox{and} \\
 (c^N_j) &=& {M_Z \over F_N} ({ s_W M_2 \over c_W^2 M_1 + s_W^2 M_2},
  -{ c_W M_1 \over c_W^2 M_1 + s_W^2 M_2}, -s_\beta{M_Z\over \mu},
   c_\beta{M_Z\over \mu}) \nonumber
\end{eqnarray}
where  $F_N=M_1 M_2 /( c_W^2 M_1 + s_W^2 M_2) + M_Z^2 s_{2\beta}/\mu$.
Here  $s_W=\sin\theta_W$ and $c_W=\cos\theta_W$ 
with the weak mixing angle $\theta_W$.

\underline{Charged lepton/chargino diagonalization}

Defining  $\theta^L$ and $\theta^R$  as the two rotation matrices
corresponding to the left-handed negatively  and positively charged 
fermions, we have
\begin{equation}
 \pmatrix{ e_i \cr \chi^-_j } \rightarrow
 \pmatrix{ e_i- \theta^L_{ik} \chi^-_k  \cr
           \chi^-_j + \theta^L_{lj} e_l } \quad;\quad
 \pmatrix{ e^c_i \cr \chi^+_j } \rightarrow
 \pmatrix{ e^c_i- \theta^R_{ik} \chi^+_k  \cr
           \chi^+_j + \theta^R_{lj} e^c_l } 
\end{equation}
where $e_i$ and $e^c_i$ denote the left-handed charged leptons and
anti-leptons, $(\chi^-_j)=(\tilde{W}^-,\tilde{H}^-_1)$ and 
$(\chi^+_j)=(\tilde{W}^+,\tilde{H}^+_2)$.
The rotation elements $\theta^{L,R}_{ij}$ are  given by
\begin{eqnarray}
&& \theta^L_{ij}= \xi_i c^L_j c_\beta-\epsilon_i \delta_{j2}\;, \quad
 \theta^R_{ij}= {m^e_i\over F_C} \xi_i c^R_j c_\beta  \quad\mbox{and} \\
&&  (c^L_j)= -{M_W \over F_C} (\sqrt{2}, 2s_\beta{M_W\over \mu})\;,
               \nonumber \\
&&   (c^R_j)= -{M_W  \over F_C} (\sqrt{2}(1-{M_2\over \mu} t_\beta), 
      \frac{M_2^2 c^{-1}_\beta}{\mu M_W }+2{M_W \over \mu} c_\beta) 
      \nonumber
\end{eqnarray}
and $F_C= M_2 + M_W^2 s_{2\beta}/\mu$.

\underline{Sneutrino/neutral Higgs boson diagonalization}

Denoting the rotation matrix by $\theta^S$, we get
\begin{equation} \label{thetaS}
 \pmatrix{ \tilde{\nu}_i \cr H^0_1 \cr H^0_2 } \rightarrow
 \pmatrix{ \tilde{\nu}_i- \theta^S_{i1} H^0_1 -\theta^S_{i2} H^{0*}_2 
                        - \theta^S_{i3} H^{0*}_1 -\theta^S_{i4} H^{0}_2  \cr
     H^0_1 + \theta^S_{i1} \tilde{\nu}_i + \theta^S_{i3} \tilde{\nu}^*_i \cr
     H^0_2 + \theta^S_{i2} \tilde{\nu}^*_i + \theta^S_{i4} \tilde{\nu}_i \cr}
\end{equation}
where
\begin{eqnarray}
 \theta^S_{i1} &=& -a_i - \eta_i s_\beta^2 m_A^2 [ m^4_{\tilde{\nu}_i} 
 - m^2_{\tilde{\nu}_i}(m_A^2+M_Z^2 s_\beta^2)-m_A^2M_Z^2s_\beta^2c_{2\beta}]
  /F_S \\
 \theta^S_{i2} &=& + \eta_i s_\beta c_\beta m_A^2 [ m^4_{\tilde{\nu}_i} 
 - m^2_{\tilde{\nu}_i}(m_A^2+M_Z^2 c_\beta^2)+m_A^2M_Z^2c_\beta^2c_{2\beta}]
  /F_S \nonumber\\
 \theta^S_{i3} &=&  -\eta_i s_\beta^2 c_\beta^2 m_A^2 M_Z^2 
       [ m^2_{\tilde{\nu}_i} - m_A^2 c_{2\beta}]/F_S \nonumber\\
 \theta^S_{i4} &=&  +\eta_i s_\beta^3 c_\beta m_A^2 M_Z^2 
       [ m^2_{\tilde{\nu}_i} + m_A^2 c_{2\beta}]/F_S 
\nonumber
\end{eqnarray}
with $F_S= (m^2_{\tilde{\nu}_i}-m_h^2)(m^2_{\tilde{\nu}_i}-m_H^2)
(m^2_{\tilde{\nu}_i}-m_A^2)$ and  $m_A, m_h$ and $m_H$ are the masses of
pseudo-scalar, light and heavy neutral scalar Higgs bosons, respectively.
Note that $m_A^2= -B\mu/c_\beta s_\beta$ in our convention.
For our calculation, we assume that all the R-parity violating parameters
are real and so are all $\theta$'s.  We also note that 
the presence of the scalar fields as well as their complex conjugates
in Eq.~(\ref{thetaS}) is due to the electroweak symmetry breaking, 
which is expected to be suppressed by the factor $M_Z^2/m_A^2$.

\underline{Charged slepton/charged Higgs boson diagonalization}

Defining $\theta^C$ as the rotation matrix, we have
\begin{equation}
 \pmatrix{ \tilde{e}_i \cr \tilde{e}^{c*}_i \cr H^-_1 \cr H^-_2 } 
 \rightarrow
 \pmatrix{ \tilde{e}_i- \theta^C_{i1} H^-_1 -\theta^C_{i2} H^{-}_2  \cr
           \tilde{e}^{c*}_i- \theta^C_{i3} H^-_1 -\theta^C_{i4} H^{-}_2  \cr
     H^-_1 + \theta^C_{i1} \tilde{e}_i + \theta^C_{i3} \tilde{e}^{c*}_i \cr
     H^-_2 + \theta^C_{i2} \tilde{e}_i + \theta^C_{i4} \tilde{e}^{c*}_i \cr}
\end{equation}
where
\begin{eqnarray}
 \theta^C_{i1} &=& -a_i - \eta_i { s_\beta^2 m_A^2 (m^2_{Ri}-m^2_{H^-})
                           \over (m^2_{H^-}-m^2_{\tilde{e}_{i1}})          
                                 (m^2_{H^-}- m^2_{\tilde{e}_{i2}}) } 
                   -\xi_i {m^e_i \mu m^2_{Di}t_\beta  
                           \over (m^2_{H^-}-m^2_{\tilde{e}_{i1}})
                                 (m^2_{H^-}- m^2_{\tilde{e}_{i2}}) }
   \\
 \theta^C_{i2} &=& - \eta_i { s_\beta c_\beta m_A^2 (m^2_{Ri}-m^2_{H^-})
                           \over (m^2_{H^-}-m^2_{\tilde{e}_{i1}})
                                 (m^2_{H^-}- m^2_{\tilde{e}_{i2}}) } 
                   -\xi_i {m^e_i \mu m^2_{Di} 
                           \over (m^2_{H^-}-m^2_{\tilde{e}_{i1}})
                                 (m^2_{H^-}- m^2_{\tilde{e}_{i2}}) } 
   \nonumber\\
 \theta^C_{i3} &=& + \eta_i { s_\beta^2 m_A^2 m^2_{Di} 
                           \over (m^2_{H^-}-m^2_{\tilde{e}_{i1}})
                                 (m^2_{H^-}- m^2_{\tilde{e}_{i2}}) } 
                   +\xi_i {m^e_i \mu  (m^2_{Li}-m^2_{H^-})t_\beta 
                           \over (m^2_{H^-}-m^2_{\tilde{e}_{i1}}) 
                                 (m^2_{H^-}- m^2_{\tilde{e}_{i2}}) }
   \nonumber\\
 \theta^C_{i4} &=& + \eta_i { s_\beta c_\beta m_A^2 m^2_{Di}
                           \over (m^2_{H^-}-m^2_{\tilde{e}_{i1}})
                                 (m^2_{H^-}- m^2_{\tilde{e}_{i2}}) } 
                   +\xi_i {m^e_i \mu (m^2_{Li}-m^2_{H^-}) 
                           \over (m^2_{H^-}-m^2_{\tilde{e}_{i1}})
                                 (m^2_{H^-}- m^2_{\tilde{e}_{i2}}) }  \,.
 \nonumber
\end{eqnarray}
Here, $m_{H^-}$ stands for the charged-Higgs boson mass, and 
$m^2_{Li}$, $m^2_{Ri}$ and $m^2_{Di}$ correspond to the LL, RR and LR
components of the $i$-th charged-slpeton mass-squared matrix, respectively, 
and $m^2_{\tilde{e}_{i1,i2}}$ are its eigenvalues. 
We remark that the appearance of $a_i$ in $\theta^S_{i1}$ and 
$\theta^C_{i1}$ is due to the rotations which remove
the Goldstone modes from the redefined neutral and charged 
slepton fields.

\medskip

With the expressions for the rotation matrices in Eqs.~(4)--(11), 
we can obtain the effective R-parity violating vertices 
from the usual R-parity conserving interaction vertices,
which are relevant to the LSP decays.
We list them below by taking only the linear terms in $\theta$'s
which are enough for our purpose.

\underline{$\chi^0-\nu-Z$ vertices}:
\begin{eqnarray} \label{chinuZ}
 {\cal L}_{\chi^0 \nu Z} &=& \overline{\chi}^0_i\gamma^\mu P_L 
         L^{\chi^0 \nu Z}_{ij} \nu_j Z_\mu^0 + h.c. 
\\ 
\mbox{with} \quad L^{\chi^0 \nu Z}_{ij} &= &{g\over 2 c_W}\,
             [c^N_1,c^N_2,0,2c^N_4]\;\xi_j c_\beta \,.
\nonumber
\end{eqnarray}

\underline{$\chi^0-l-W$ vertices}:
\begin{eqnarray} \label{chilW}
 {\cal L}_{\chi^0 l W} &=& \overline{\chi}^0_i\gamma^\mu 
  \left[ P_L L^{\chi^0 l W}_{ij}  + P_R R^{\chi^0 l W}_{ij} \right]
          e_j W_\mu^+ + h.c. 
\\ 
\mbox{with} \quad L^{\chi^0 l W}_{ij} &= &{g\over \sqrt{2}} \,
             [c^N_1,c^N_2-\sqrt{2}c^L_1, c^N_3-c^L_2,c^N_4] \,
             \xi_j c_\beta \nonumber \\
           \quad R^{\chi^0 l W}_{ij} &= &{g\over \sqrt{2}} \,
             [0,-\sqrt{2}c^R_1, 0, -c^R_2] \,
             \xi_j c_\beta 
\nonumber
\end{eqnarray}


\underline{$\chi^0-\nu-H^0_{1,2}$ vertices}:
\begin{eqnarray}
 {\cal L}_{\chi^0 \nu H^0_{1,2}} &=& \overline{\chi}^0_i 
  \left[ P_L L^{\chi^0 \nu H^0_{1,2}}_{ij}
         + P_R R^{\chi^0 \nu H^0_{1,2}}_{ij} \right]
          \nu_j H^{0*}_{1,2} + h.c. 
\\
\mbox{with} \quad 
  L^{\chi^0 \nu H^0_{1}}_{ij} &= &{g\over \sqrt{2}} \,
             [-t_W(\theta^S_{j1}-\theta^N_{j3}), 
                ( \theta^S_{j1}-\theta^N_{j3}), 
                 (t_W \theta^N_{j1}-\theta^N_{j2}), 0 ] 
             \nonumber \\
  L^{\chi^0 \nu H^0_{2}}_{ij} &= &{g\over \sqrt{2}} \,
             [-t_W(\theta^S_{j4}+\theta^N_{j4}), 
                 (\theta^S_{j4}+\theta^N_{j4}),  0, 
                 (-t_W\theta^N_{j1}+\theta^N_{j2}) ] 
             \nonumber \\
  R^{\chi^0 \nu H^0_{1}}_{ij} &= &{g\over \sqrt{2}} \,
           [-t_W \theta^S_{j3}, \theta^S_{j3}, 0, 0 ]
             \nonumber\\
  R^{\chi^0 \nu H^0_{2}}_{ij} &= &{g\over \sqrt{2}} \,
           [-t_W \theta^S_{j2}, \theta^S_{j2}, 0, 0 ]
\nonumber
\end{eqnarray}

\underline{$\chi^0-l-H^+_{1,2}$ vertices}:
\begin{eqnarray}
 {\cal L}_{\chi^0 l H^+_{1,2}} &=& \overline{\chi}^0_i 
  \left[ P_L L^{\chi^0 l H^+_{1,2}}_{ij}
         + P_R R^{\chi^0 l H^+_{1,2}}_{ij} \right]
          e_j H^{+}_{1,2} + h.c. 
\\ 
\mbox{with} \quad 
  L^{\chi^0 l H^+_{1}}_{ij} &= &{-1\over \sqrt{2}} \,
             [g'(\theta^C_{j1}-\theta^L_{j2}), 
               g(  \theta^C_{j1}-\theta^L_{j2}), 
                 \sqrt{2}(g\theta^L_{j1}+h^e_j\theta^C_{j3}), 0 ] 
             \nonumber \\
  L^{\chi^0 l H^+_{2}}_{ij} &= &{-1\over \sqrt{2}} \,
             [g' \theta^C_{j2},  g \theta^C_{j2},
                 h^e_j\theta^C_{j4}, 0 ] 
             \nonumber \\
  R^{\chi^0 l H^+_{1}}_{ij} &= &
           [\sqrt{2} g' \theta^C_{j3} + h^e_j \theta^N_{j1}, 
            h^e_j \theta^N_{j2}, 
            -h^e_j (\theta^C_{j1}-\theta^N_{j3}), 
            h^e_j \theta^N_{j4} ]
             \nonumber\\
  R^{\chi^0 l H^+_{2}}_{ij} &= &
           [\sqrt{2} g' \theta^C_{j4} - {g'\over\sqrt{2}} \theta^R_{j2}, 
            - {g\over\sqrt{2}} \theta^R_{j2}, 
            - h^e_j \theta^C_{j2}, 
            - g \theta^R_{j1} ]
\nonumber
\end{eqnarray}


%
%

In Eqs.~(12)--(15), 
the four components inside brackets correspond to
the indices $i=1,\cdots,4$ indicating the neutralino states
($\tilde{B}$, $\tilde{W}_3$, $\tilde{H}^0_1$, $\tilde{H}^0_2$), 
respectively,  as before.
Here, let us remark that all of the above vertices  depend
only on the variables $\xi_i$ or $\eta_i$ which are generated by 
renormalization group evolution under the universality condition,  
even though
the individual elements $\theta^N_{i3}$, $\theta^L_{i2}$, 
$\theta^S_{i1}$ and $\theta^C_{i1}$ depend on either $\epsilon_i$ 
or $a_i$.  This fact will be important when we study the LSP decay 
processes.

\medskip

\underline{Effective $LQ\bar{d}$, $L\bar{Q}u$, $LL\bar{e}$
and $\nu f \tilde{f}^*$  vertices}:

In the below, we list the $\lambda$-like or $\lambda'$-like couplings 
which are, however, neither supersymmetric nor $SU(2)_L$-symmetric:
\begin{eqnarray}
&& {\cal L}_{LQ\bar{d}} = \varepsilon_{ab}\left[
  {\Lambda}^{d1}_{aij} \tilde{L}_{ai}\overline{d}_j P_L Q_{bj}
  + {\Lambda}^{d2}_{aij}\tilde{L}^{c*}_{ai}\overline{d}_j P_L Q_{bj}
    \right. 
\\
&& \qquad\qquad \left. 
   + \Lambda^{d3}_{aij} \left( \overline{d}_j P_L L_{ai} \tilde{Q}_{bj}
    +\overline{L}^c_{ai} P_L Q_{bj} \tilde{d}^c_j \right) 
    +\Lambda^{d4}_{ai} \overline{L}^c_{ai} P_L Q_{bj} \tilde{d}^*_j 
     \right] + h.c.
  \nonumber\\
\mbox{where}&&
   {\Lambda}^{d1}_{aij} = 
          [\theta^S_{i1},\theta^C_{i1}]\, h^d_j \;, \quad
   {\Lambda}^{d2}_{aij} = 
          [\theta^S_{i3},\theta^C_{i3}]\, h^d_j \;,
 \nonumber\\
&&   \Lambda^{d3}_{aij} = 
          [\theta^N_{i3},\theta^L_{i2}]\, h^d_j\;, \quad
  \Lambda^{d4}_{ai} = {g\over\sqrt{2}}
          [-t_W \theta^N_{i1}+\theta^N_{i2}, \sqrt{2}\theta^L_{i1}]
 \nonumber
\end{eqnarray}

\begin{eqnarray}
&& {\cal L}_{L\bar{Q}u} = \delta_{ab}\left[
    {\Lambda}^{u1}_{aij} \tilde{L}_{ai}\overline{Q}_{bj} P_R u_{j}
  + {\Lambda}^{u2}_{aij}\tilde{L}^{c*}_{ai}\overline{Q}_{bj} P_R u_{j}
    \right. 
\\
&& \qquad\qquad \left. 
  + {\Lambda}^{u3}_{aij} \left(  
      \tilde{u}^{c*}_{j}\overline{Q}_{bj} P_R L_{ai}
       +\overline{L}^c_{ai}  \tilde{Q}^*_{bj} P_R u_j \right) 
    +{\Lambda}^{u4}_{aij} \tilde{u}_j \overline{Q}_{aj} P_R L_{ai} 
     \right] + h.c.
  \nonumber\\
\mbox{where} &&
   {\Lambda}^{u1}_{aij} = 
          [-\theta^S_{i2},\theta^C_{i2}]\, h^u_j \;, \quad
   {\Lambda}^{u2}_{aij} = 
          [-\theta^S_{i4},\theta^C_{i4}]\, h^u_j \;,
 \nonumber\\
&&   \Lambda^{u3}_{aij} = 
          [-\theta^N_{i4},\theta^R_{i2}]\, h^u_j \;, \quad
 {\Lambda}^{u4}_{ai} = {g\over\sqrt{2}}
          [-{1\over3}t_W \theta^N_{i1}-\theta^N_{i2}, -\sqrt{2}\theta^R_{i1}]
\nonumber
\end{eqnarray}

\begin{eqnarray}
&& {\cal L}_{LL\bar{e}} = \varepsilon_{ab}\left[
  {\Lambda}^{l1}_{aij} \tilde{L}_{ai}\overline{e}_j P_L L_{bj}
  + {\Lambda}^{l2}_{aij}\tilde{L}^{c*}_{ai}\overline{e}_j P_L L_{bj}
    \right. 
\\
&& \qquad\qquad \left. 
   + \Lambda^{l3}_{aij} \left( \overline{e}_j P_L L_{ai} \tilde{L}_{bj}
    -\overline{L}^c_{ai} P_L L_{bj} \tilde{e}^c_j \right) 
    +\Lambda^{l4}_{ai} \overline{L}^c_{ai} P_L L_{bj} \tilde{e}^*_j 
     \right]  
\nonumber\\
&& \qquad\qquad
  +\Lambda^{l5}_{ai} \tilde{\nu}_j \overline{L}^c_{aj} P_R L_{ai} + h.c. 
\nonumber\\
\mbox{where} &&
   {\Lambda}^{l1}_{aij} = 
          [\theta^S_{i1},\theta^C_{i1}]\, h^e_j \;,\quad
   {\Lambda}^{l2}_{aij} = 
          [\theta^S_{i3},\theta^C_{i3}]\, h^e_j \;,\quad
   \Lambda^{l3}_{aij} = 
          [\theta^N_{i3},\theta^L_{i2}]\, h^e_j \nonumber  \\
&&   \Lambda^{l4}_{ai} = {g\over\sqrt{2}}
          [t_W \theta^N_{i1}+\theta^N_{i2}, -\sqrt{2}\theta^L_{i1}] 
     \,,\quad
     \Lambda^{l5}_{ai} = {g\over\sqrt{2}}
          [t_W \theta^N_{i1}-\theta^N_{i2}, -\sqrt{2}\theta^R_{i1}]
\nonumber
\end{eqnarray}
In Eqs.~(16)--(18), the two components in the brackets correspond to
the two states of the $SU(2)_L$ doublets with indices $a,b=1,2$,  and 
$L^c \equiv (\nu, e^c)$ is defined as an lepton $SU(2)_L$ doublet while
$\tilde{L}^c=(\tilde{\nu},\tilde{e}^c)$ is its scalar counterpart.

Finally, we have
\begin{equation}
{\cal L}_{\nu f \tilde{f}} =
 \Lambda^{\nu}_i\, \overline{\nu}_i P_R \left[ 
 {2\over3} u_j \tilde{u}^c_j -{1\over3} d_j \tilde{d}^c_j
 - e_j \tilde{e}^c_j \right] + h.c.
\end{equation}
where $ \Lambda^{\nu}_i = \sqrt{2} g' \theta^N_{i1} $.

As one can see, the above vertices are non-supersymmetric and
$SU(2)_L$ breaking.  But, among various terms in Eqs.~(16) and (18),
one can separate out the supersymmetric couplings,
$\epsilon_i h^d_j$ and $\epsilon_i h^e_j$, leaving 
all the vertices depending only on $\xi_i$ or $\eta_i$ similarly to 
the vertices in Eqs.~(12)--(15).   Then, combining those with
the couplings in the superpotential (\ref{supo}), we can define
the effective supersymmetric couplings as
$\tilde{\lambda}'_{ijk} = \epsilon_i h^d_j \delta_{jk}
+\lambda'_{ijk}$ and 
$\tilde{\lambda}_{ijk} = \epsilon_i h^e_j \delta_{jk}
+\lambda_{ijk}$.  We will see that 
these couplings determine the quantity $\xi_i$ or $\eta_i$ through
the renormalization group evolution of the bilinear (soft) terms.

\section{Radiative neutrino mass matrix from R-parity violation}

After performing the rotations described in the previous section, 
the three neutrinos in the ``weak-basis'' get important mass corrections
arising from the see-saw mechanism associated with the heavy four 
neutralinos.  As is well-known,  this gives the ``tree-level'' neutrino 
matrix of the form;
\begin{equation} \label{Mtree}
 M^{tree}_{ij}= - {M_Z^2 \over F_N} \xi_i \xi_j c_\beta^2 \,,
\end{equation}
which makes massive only one neutrino, $\nu_3$, 
in the direction of $\vec{\xi}$.  The other two get masses from
finite one-loop corrections and thus $\nu_3$ is usually 
the heaviest component.  We fix the value of 
$m_{\nu_3}$  from the atmospheric neutrino data \cite{sk-atm} 
and thus the overall size of 
$\xi\equiv |\vec{\xi}|$ as
\begin{equation} \label{xicb}
 \xi c_\beta = 0.74\times10^{-6} \left(F_N \over M_Z\right)^{1/2}
               \left(m_{\nu_3} \over 0.05 \mbox{ eV} \right)^{1/2}\,.
\end{equation}
where $F_N$ is defined in Eq.~(5) and its typical value is given by
$M_2$.
Furthermore, among three neutrino mixing angles defined by
the mixing matrix 
\begin{equation}
 U=\pmatrix{ 1 & 0 & 0 \cr 0 & c_{23} & s_{23} \cr 0 & -s_{23} & c_{23} \cr}
   \pmatrix{ c_{13} & 0 & s_{13} \cr 0 & 1 & 0 \cr -s_{13} & 0 & c_{13} \cr}
   \pmatrix{ c_{12} & s_{12} & 0 \cr -s_{12} & c_{12} & 0 \cr 0 & 0 & 1 \cr}
\end{equation}
with $c_{ij}=\cos\theta_{ij}$ and $s_{ij}=\sin\theta_{ij}$, etc.,
two angles are almost determined by the tree-level mass matrix (\ref{Mtree})
as follows;
\begin{eqnarray} \label{twoangles}
  \sin^22\theta_{atm} &\approx&
  \sin^22\theta_{23} \approx 4 {\xi_2^2 \over \xi^2}
                                 {\xi_3^2\over \xi^2} \nonumber\\
  \sin^22\theta_{chooz} &\approx&
  \sin^22\theta_{13} \approx 4 {\xi_1^2 \over \xi^2}
                         \left(1-{\xi_1^2\over \xi^2}\right)  \,.
\end{eqnarray}
The atmospheric neutrino and CHOOZ  experiments  \cite{sk-atm,chooz}
require 
$\sin^22\theta_{atm} \approx 1$ and $\sin^22\theta_{chooz} < 0.2$.
The angle $\theta_{12}$ can be determined only after including
one-loop corrections and is responsible for the solar neutrino mixing,
$\theta_{sol} \approx \theta_{12}$,
if the neutrino mass matrix is to explain the atmospheric and solar
neutrino oscillations as will be discussed in the next section.

\medskip

An important property of the tree-level neutrino mass matrix (\ref{Mtree}) 
is that it  depends on the bilinear R-parity violating quantities 
$\xi_i=a_i - \epsilon_i$ which can be re-expressed as
\begin{equation} \label{xiis}
 \xi_i = \epsilon_i {\Delta m^2_i+ \Delta B_i \mu t_\beta \over
                     m_{\tilde{\nu}_i}^2 }
         - {m^2_{L_i H_1} \over m_{\tilde{\nu}_i}^2 }
\end{equation}
where $\Delta m^2_i \equiv m^2_{H_1}-m^2_{L_i}$,
$\Delta B_i \equiv B-B_i$ and $ m_{\tilde{\nu}_i}^2 =
m^2_{L_i}+ M_Z^2 c_{2\beta}/2$.  In other words, the nonzero values
of $\xi_i$ arise from the mismatch of soft mass parameters for
the Higgs field $H_1$ and slepton field $L_i$ having the same gauge
quantum numbers.  If one assumes the universality condition,
one has $\Delta m^2_i= \Delta B_i = m^2_{L_i H_1}=0$ 
at the mediation scale of supersymmetry breaking,  and
their nonzero values are generated by Yukawa coupling effects 
through the renormalization group  evolution down to the weak scale.

Under the assumption that the R-parity violating  couplings
follow the usual hierarchies as the quark and lepton Yukawa
couplings, that is, $\lambda'_i\equiv\lambda'_{i33}$ and 
$\lambda_i\equiv\lambda_{i33}$ give the 
dominant contributions, the renormalization group equations (RGE) 
of the bilinear terms are given by 
\begin{eqnarray} \label{rges}
 16 \pi^2 {d\over dt} \Delta m^2_i &=& 
      6 h_b^2 X_b + 2 (1-\delta_{i3}) h_\tau^2 X_\tau \nonumber\\
 16 \pi^2 \epsilon_i {d\over dt} \Delta B_i &=& 
      \epsilon_i (6 h_b^2 A_b + 2(1-\delta_{i3}) h_\tau^2 A_\tau )
      + (6 \lambda'_i h_b A'_i + 2\lambda_i h_\tau A_i)   \nonumber\\
 &&     - \Delta B_i (3 \lambda'_i h_b  + \lambda_i h_\tau)  \\
 16 \pi^2 {d\over dt} m^2_{L_i H_1} &=& 
      -(6 \lambda'_i h_b X_b +  2\lambda_i h_\tau X_\tau)
   +  m^2_{L_i H_1} (3 h_b^2 +(1+\delta_{i3}) h_\tau^2) \nonumber\\
&&  +(6 \lambda'_i h_b +  2\lambda_i h_\tau )\Delta m^2_i
   -(6 \lambda'_i h_b A_b \Delta A'_i +  2\lambda_i h_\tau A_\tau
           \Delta A_i)           \nonumber
\end{eqnarray}
where $t=\ln Q$ with the renormalization scale $Q$, 
$X_b=m^2_{Q_3}+m^2_{D^c_3} + m^2_{H_1} + A_b^2$, 
$X_\tau=m^2_{L_3}+m^2_{E^c_3} + m^2_{H_1} + A_\tau^2$.
Here, $A$'s are the trilinear soft parameters corresponding to
the $h_b,h_\tau$, $\lambda'_i$ and $\lambda_i$ couplings, and finally
$\Delta A'_i\equiv A'_i-A_b$, $\Delta A_i\equiv A_i-A_\tau$. 
Under the one-step approximation, the above RGE can be solved as
\begin{eqnarray} \label{rgesol}
 \epsilon_i \Delta m^2_i - m^2_{L_i H_1} &=& 
     {1 \over 8\pi^2} \left(3 \tilde{\lambda}'_i h_b X_b 
             +  \tilde{\lambda}_i h_\tau X_\tau \right)
         \ln{M_m \over m_{\tilde{t}} }
        \nonumber\\
 \epsilon_i  \Delta B_i &=& 
     {1 \over 8\pi^2} \left(3 \tilde{\lambda}'_i h_b \tilde{A}'_i 
             +  \tilde{\lambda}_i h_\tau \tilde{A}_i \right)
         \ln{M_m \over m_{\tilde{t}} }
\end{eqnarray}
where 
$M_m$ is the mediation scale of supersymmetry breaking,  and 
$m_{\tilde{t}}$ is a typical stop mass scale where we 
calculate the sneutrino vacuum expectation values.
Here, 
we have defined $\tilde{\lambda}'_i = \epsilon_i h_b + \lambda'_i$,
$\tilde{\lambda}_i = \epsilon_i(1-\delta_{i3}) h_\tau  + \lambda_i$, 
$\tilde{\lambda}'_i \tilde{A}'_i  = \epsilon_i h_b A_b +  \lambda'_i A'_i$ 
and 
$\tilde{\lambda}_i \tilde{A}_i  = \epsilon_i(1-\delta_{i3}) h_\tau A_\tau 
+  \lambda_i A_i$. 
Note that $\lambda_3$ as well as $\tilde{\lambda}_3$ vanish.
In gauge mediated supersymmetry breaking models where
the mediation scale $M_m$ is low,  the above approximate solution is 
quite reliable.   

\medskip

We are ready to discuss
the typical sizes of the supersymmetric bilinear,
and trilinear parameters, $\epsilon_i$ and $\lambda'_i, \lambda_i$,
(or $\tilde{\lambda}'_i$, $\tilde{\lambda}_i$)
which will be relevant for the study of the LSP decay.
Assuming that there is no fine cancellation among various terms 
in Eq.~(\ref{xiis}) and the term with $X_b$ gives the largest contribution
in (\ref{rgesol}), we obtain 
\begin{equation}
 {\xi_i c_\beta \over \tilde{\lambda}'_i} \sim {3\over 8 \pi^2} 
        {m_b \over v}{ X_b \over m^2_{\tilde{\nu}_i} } 
         \ln{ M_m  \over m_{\tilde{t}} }  \,.
\end{equation}
In gauge mediated supersymmetry breaking models \cite{gmsb},
the sfermion soft masses are determined by gauge-boson/gaugino loop 
corrections which implies  $X_b/m^2_{\tilde{\nu}_i}
\approx 2\alpha_3^2/\alpha_2^2$.   Further assuming the supersymmetry 
breaking scale  $\Lambda_S$ close to $M_m$, we take  
$M_m/m_{\tilde{t}} \approx 4\pi/\alpha_3$.  This gives  
\begin{equation} \label{lampv}
\epsilon_i h_b\;\mbox{or}\; \lambda'_i \sim 20\, \xi_i c_\beta \,.
\end{equation}
When the tree mass matrix gives the atmospheric
neutrino mass scale as discussed,  we get $\xi_i c_\beta \sim 10^{-6}$
and thus
\begin{equation}
 \epsilon_i \sim 20 \xi_i c_\beta/h_b \sim 2\times10^{-3}/t_\beta
\end{equation}
for $F_N=M_Z$.  
This shows that the parameters $\epsilon_i$ and $a_i$ can be very large
while maintaining $\xi_i=a_i-\epsilon_i$ very small for low $\tan\beta$.

Let us consider another possibility that 
$\tilde{\lambda}_i$ gives dominant contribution in Eq.~(\ref{rgesol}).
Following the similar steps as above, we obtain
\begin{equation} \label{lambv}
  \epsilon_i h_\tau\;\mbox{or}\; \lambda_i \sim 560\, \xi_i c_\beta
\end{equation}
where we took $X_\tau = 3 m^2_{\tilde{\nu}_i}$.
This implies that the contribution of $\lambda_i$ to $\xi_i$ is
comparable to that of $\lambda'_i$ if $\lambda_i \sim 30 \lambda'_i$.
Later, we will see that the large mixing angle explaining the
the solar neutrino data
can be obtained for $\lambda_{1,2} \sim 5 \lambda'_{2,3}$.

\medskip

So far, we neglected the radiative corrections in the determination 
of vacuum expectation values of the sneutrino as well as Higgs fields. 
To obtain reliable minimization conditions for the electroweak 
symmetry breaking, one has to consider the effective scalar potential
$$ V_{eff}= V_0 + V_1$$
where $V_0$ is the tree-level potential and $V_1=
{1\over 64\pi^2} \mbox{Str} {\cal M}^4 
\left( \ln{{\cal M}^2 \over Q^2} -{3\over2}\right)$
includes one-loop corrections.  With R-parity violation, $V_1$ is a function
of not only the Higgs fields but also the sneutrinos \cite{CK,valle97,valle1}.
In deriving Eq.~(\ref{xiis}), we neglected $V_1$ and used the tree-level
minimization conditions for the neutral Higgs and sneutrino fields.  
Since the nonzero $\xi_i$'s  are also generated by the renormalization
effect, the inclusion of $V_1$ in the determination of sneutrino 
vacuum expectation values is crucial, in particular, in the case of 
a low-scale  supersymmetry breaking mediation. 
In gauge mediation models,  such one-loop 
corrections can give rise to an order of magnitude change
in neutrino mass-squared values which are well-measured in the 
atmospheric and solar neutrino experiments \cite{CK}.
After including such effects, Eq.~(\ref{xiis}) is modified to 
\begin{eqnarray} \label{newxi}
 \xi_i &=& + \epsilon_i {\Delta m^2_i+ \Delta B_i \mu t_\beta \over
                     m_{\tilde{\nu}_i}^2 +\Sigma^{(2)}_{L_i}}
         - {m^2_{L_i H_1} \over m_{\tilde{\nu}_i}^2+\Sigma^{(2)}_{L_i} }
     \nonumber\\
& &  +  \epsilon_i {\Sigma_{H_1} - \Sigma^{(2)}_{L_i}
       - \epsilon_i^{-1}\Sigma^{(1)}_{L_i} 
      \over m_{\tilde{\nu}_i}^2+\Sigma^{(2)}_{L_i} }
\end{eqnarray}
where $\Sigma_{H_1} = \partial V_1/H_1^* \partial H_1$, 
$\Sigma^{(1)}_{L_i} = \partial V_1/H_1^* \partial L_i$  and 
$\Sigma^{(2)}_{L_i} = \partial V_1/L_i^* \partial L_i$ \cite{CK}.
In our numerical calculation in the following section, we
include such improvements.

\medskip

In order to get the full neutrino mass matrix, one-loop radiative
corrections to neutrino mass matrix should be included:  
$$ M^\nu_{ij} = M^{tree}_{ij} + M^{loop}_{ij} \,. $$
In the below, we will discuss whether the above mass matrix
$M^\nu$ can explain the atmospheric and solar neutrino data, 
simultaneously.  The numerical calculation in this direction
has been performed first in Ref.~\cite{hemp} without including 
the effect of the one-loop effective potential $V_1$ in the
context of minimal supergravity models.  Inclusion of full one-loop
corrections has been made in Refs.~\cite{CK} and \cite{valle1}, in
gauge mediation and supergravity models, respectively.
As mentioned, the one-loop mass matrix  $M^{loop}$  lifts twofold
degeneracy of the tree-level mass matrix and thus all the three neutrinos
get masses which are generically hierarchical.
It is instructive to compare the tree and loop mass components,
in order to get an idea about the mass of the neutrino, $\nu_2$, which
determine the solar neutrino mass scale.
The largest contribution to $M^{loop}$ usually comes from 
the one-loop diagrams with $\lambda'_i$ and  $\lambda_i$ (more generally 
with the induced ones, $\tilde{\lambda}'_i$ and  $\tilde{\lambda}_i$)
which takes the from
\begin{eqnarray} \label{Mloop}
M^{loop}_{ij} &=& 3 {\tilde{\lambda}'_i \tilde{\lambda}'_j \over 8\pi^2}
   {m_b^2(A_b+\mu t_\beta) \over m^2_{\tilde{b}_1} - m^2_{\tilde{b}_2} }
     \ln{m^2_{\tilde{b}_1} \over m^2_{\tilde{b}_2}}
        \nonumber\\
           && + { \tilde{\lambda}_i \tilde{\lambda}_j \over 8\pi^2}
   {m_\tau^2(A_\tau+\mu t_\beta) \over 
     m^2_{\tilde{\tau}_1} - m^2_{\tilde{\tau}_2} }
     \ln{m^2_{\tilde{\tau}_1} \over m^2_{\tilde{\tau}_2}}
\end{eqnarray}
where $m_{\tilde{b}_i}$ and $m_{\tilde{\tau}_i}$ are the
sbottom and stau mass eigenvalues, respectively.

When $\tilde{\lambda}'_i > \tilde{\lambda}_i$ so that  
$\tilde{\lambda}'_i$ give dominant contributions to  both $\xi_i$ and
$M^{loop}$,  one has
\begin{equation} \label{bicase}
 {M^{loop}_{ij} \over M^{tree}_{ij} }
 \sim {3\over 8 \pi^2} {\tilde{\lambda}'_i \tilde{\lambda}'_j 
      \over  \xi_i \xi_j c_\beta^2}
       {m_b^2 \mu t_\beta \over M_Z m^2_{\tilde{b}}} \sim 10^{-3} t_\beta
\end{equation}
assuming the relation (\ref{lampv}) and $2.5\mu=m_{\tilde{b}}=500$ GeV.
In this case, the second neutrino mass eigenvalue is determined by
the sub-leading contribution of $\tilde{\lambda}_i$ to either
$\xi_i$ or $M^{loop}$.  Thus, we expect
$ m_{\nu_2}/m_{\nu_3} < 10^{-3} t_\beta$, or equivalently,
\begin{equation} \label{smallratio}
 {\Delta m^2_{21} \over \Delta m^2_{32} }
 < 10^{-6} t_\beta^2 \,.
\end{equation}
Note that the solar neutrino experiments  require $\Delta m^2_{21}
= 10^{-5}-10^{-10}$ eV$^2$ depending on the type of solar 
neutrino oscillation solutions.  
As the atmospheric neutrino oscillation requires
$\Delta m^2_{32}\approx 3\times10^{-3}$ eV$^2$, it would be
much easier to get the so-called vacuum oscillation or
the low $\Delta m^2$ MSW solution.  To realize the large mixing MSW
(LMA) solution which is now strongly favored by the recent SNO data 
\cite{sol-exp},
a large $\tan\beta$ is needed.  Such a tendency has also been  
observed by numerical calculations in the context of minimal
supergravity models \cite{hemp,valle1}.  However, under the assumption
that $\tilde{\lambda}'_i > \tilde{\lambda}_i$, it is impossible to get a
large mixing angle for the solar neutrino oscillation due to the 
CHOOZ constraint.  This will become clear when we 
discuss the bilinear model in the next section.

Let us now consider the opposite case that 
$\tilde{\lambda}_i \gg \tilde{\lambda}'_i$  
so that $\tilde{\lambda}_i$ give dominant contributions to 
$\xi_i$ and $M^{loop}$
(which is the case when  $\tilde{\lambda}_i > 30 \tilde{\lambda}'_i$
as in Eq.~(\ref{lambv})), one finds
\begin{equation} \label{tricase}
 {M^{loop}_{ij} \over M^{tree}_{ij} }
 \sim {1\over 8 \pi^2} {\tilde{\lambda}_i \tilde{\lambda}_j
      \over  \xi_i \xi_j c_\beta^2}
       {m_\tau^2 \mu t_\beta \over M_Z m^2_{\tilde{\tau}}} \sim t_\beta
\end{equation}
for $\mu=1.5 m_{\tilde{\tau}}=200$ GeV.  Note that 
$m_{\nu_2}/m_{\nu_3}$ can be even larger than one
and the resultant neutrino mass components satisfy
$M^{\nu}_{11,12,22} > M^{\nu}_{i3}$ as $\tilde{\lambda}_3\equiv0$.
Such a case is not favorable as it cannot give a large mixing angle 
for the the atmospheric neutrino oscillation.
From the above discussion, we can infer that the atmospheric and 
LMA solar neutrino oscillation  can be realized if
the couplings $\tilde{\lambda}_i$ and $\tilde{\lambda}'_i$
satisfy a relation in-between (33) and (35), which will be the
case that $\lambda_i$ is moderately larger than $\lambda'_i$.\footnote{
The LMA solution may also be obtained  in the bilinear 
model if one relaxes the universality condition \cite{valle1,anjan}.}
We will analyze the neutrino masses and mixing in such a scheme
and its collider signature in the following section.

\section{Atmospheric and solar neutrino oscillations and
LSP decays in GMSB models}

Let us make a numerical analysis to find how the neutrino mass matrix 
from R-parity violation explain both the atmospheric 
and solar neutrino oscillations 
and what are the corresponding collider signatures coming
from the LSP decay.  Our discussions are  specialized in
the models with gauge mediated supersymmetry breaking (GMSB)
in which the universality condition is automatic and thus
supersymmetric flavor problems are naturally avoided.
For our discussion, we will take the minimal number of the
messenger multiplets ($5+\bar{5}$) and the messenger supersymmetry
breaking scale $\Lambda_S$ not too far from the mediation scale $M_m$
\cite{gmsb}.  We will also concentrate on the cases
where the LSP (being a neutralino) is lighter than the $W$ boson 
so that only three-body decays are allowed.

Due to the (effective) R-parity violating couplings introduced 
in the previous sections, the LSP, denoted by $\tilde{\chi}^0_1$,
decays through the mediation of on/off-shell $W,Z$ gauge bosons, Higgses, 
sleptons and squarks, producing the  following three-fermion final states,
$$ \nu \nu  \nu\,, \quad    \nu l_i^\pm l_j^\mp\,,\quad
   \nu q \bar{q}'\,,\quad   \l_i^\pm q q' \,.  $$
Here we do not distinguish the neutrino flavors, and 
the final quark states will be identified with jets.
The modes,  $\nu l^\pm_i l^\mp_j$ and $\l^\pm_i jj$, are of a
particular interest since the flavor dependence of R-parity 
violating couplings, which are relevant to the neutrino mixing angles,
will be encoded in their branching ratios.

\medskip

Before performing the  numerical analysis, let us make some
qualitative discussions on the LSP decay.
When the LSP is heavier than the $W$ boson, the decay modes 
$\tilde{\chi}^0_1 \to l_i^\pm W^\mp$ will have  sizable branching fractions
\cite{mrv,jaja} and measuring them will give a direct
information on the ratios $ \xi_1^2 : \xi_2^2 : \xi_3^2$ from which we can
probe the neutrino mixing angles $\theta_{23}$ and $\theta_{13}$ through
Eq.~(\ref{twoangles}) \cite{jaja}.  
The decay rate of the mode $\tilde{\chi}^0_1 \to
l_i W$ is given by \cite{jaja}
\begin{eqnarray} \label{lW}
\Gamma(l_i W) &=& {G_F m^3_{\tilde{\chi}^0_1} \over 4\sqrt{2} \pi} 
   [|C^L_1|^2 + |C^R_1{m^e_i\over F_C}|^2] |\xi_i|^2 c_\beta^2\, 
    I_2(M_W^2/m^2_{\tilde{\chi}^0_1})
\\
\mbox{with} \quad
 C^L_1 &=& {1\over\sqrt{2}}  [N_{11} c^N_1 + N_{12}(c^N_2-\sqrt{2}c^L_1)
+N_{13}(c^N_3-c^L_2)+N_{14}c^N_4]
 \nonumber\\
 C^R_1 &=& N_{12}c^R_1 +  {1\over \sqrt{2}}  N_{14}c^R_2.
\nonumber
\end{eqnarray}
Here, $N_{1j}$ are the components of the
neutralino diagonalization matrix for the LSP
and $I_2(x)=(1-x)^2(1+2x)$. 
Taking  $\xi_ic_\beta=10^{-6}$, $C_1^L=1$ and $m_{\tilde{\chi}^0_1}=M_Z$, 
we get $\Gamma(lW) \approx 10^{-14}$ GeV corresponding to the  
decay length $\tau \approx 2\,cm$.
Thus, the measurement of BR($e W$) : BR($\mu W$) : BR($\tau W$), or
$\xi_1^2$ : $\xi_2^2$ : $\xi_3^2$,  will be certainly feasible 
in the future colliders.
Note that the contribution $C_1^R m_\tau/F_C \sim m_\tau t_\beta/\mu$ 
can be neglected unless $\tan\beta$ is very large.

If the LSP is lighter than the $W$ boson,
only three-body decay modes are allowed and thus the desired 
decay modes may be too suppressed to be observed.
As a comparison with the above two body decay, let us consider the 
process $\tilde{\chi}^0_1 \to l_i W^* \to l_i f f'$ whose decay rate is
\begin{equation} \label{lW*}
 \Gamma(l_i W^*) = {3 G_F^2 m_{\tilde{\chi}^0_1}^5 \over 64\pi^3} 
           [|C_1^L|^2+|C_1^R {m^e_i \over F_C}|^2] |\xi_i|^2 c^2_\beta
          \, I_3(m_{\tilde{\chi}^0_1}^2/M_W^2)
\end{equation}
where $I_3(x)=[12x-6x^2-2x^3+12(1-x)\ln(1-x)]/x^4$.
This gives $\Gamma(l_i W^*) \approx 8\times10^{-17}$ GeV for 
$C_1^L=1$, $\xi_i c_\beta= 10^{-6}$ and $m_{\tilde{\chi}^0_1}=50$ GeV.
If this is the dominant decay channel, the total decay length will be
$\tau \sim 2.5 m$ making it hard to observe sufficient LSP decay signals.
However, it will turn out that the dominant LSP decay diagrams involve the
effective $\lambda'_i$ or $\lambda_i$ couplings which make the LSP
decay well inside the detector.
This can be understood from the previous discussions 
showing that $\tilde{\lambda}'_i, \tilde{\lambda}_i
\gg \xi_i c_\beta$.  Furthermore, the corresponding decay modes
$\tilde{\chi}^0_1 \to \nu jj$ or $\nu l_i^\pm l_j^\mp$ are 
dominated by the diagrams with the exchange of the sneutrino or 
charged slepton (in particular, the right-handed stau) which are 
relatively light.  To get an order of magnitude estimation,
let us consider the decay rate for 
$\tilde{\chi}_1^0 \to \nu l_i^\pm l_j^\mp$;
\begin{equation}
\Gamma(\nu_k l_i^\pm l_j^\mp) =
 {\alpha' \tilde{\lambda}_{kij}^2 \over 768 \pi^2}  
 {  m_{\tilde{\chi}_1^0}^5 \over m^4_{\tilde{e}^c_j} }
  |N_{11}|^2  J(m^2_{\tilde{\chi}_1^0}, m^2_{\tilde{\nu}_i},
          m^2_{\tilde{e}_j}, m^2_{\tilde{e}^c_k})
\end{equation}
where $\alpha'=g'^2/4\pi$ and 
$J$ is a order-one function of the sparticle masses
which is normalized to be one in the limit of 
$m^2_{\tilde{\chi}_1^0}= m^2_{\tilde{e}^c_k}=0$.
Taking $\lambda_{i33}=2\times10^{-5}$, $m_{\tilde{\chi}_1^0}=50$ GeV
$m_{\tilde{e}^c_3}=70$ GeV and $J=1$, we get
$\Gamma(\nu_i \tau\bar{\tau}) \approx \Gamma(\nu_3 l_i^\pm \tau^\mp)
\approx10^{-14}$ GeV which corresponds to $\tau \sim 2\, cm$.  
As we will see, this is a typical order of magnitude for  the
total decay rate of the LSP when R-parity violation accounts for
the atmospheric and solar neutrino masses and mixing.

\medskip

Let us now present our numerical results  for the
two possible schemes of R-parity violation:
(i) the bilinear model which has only three input parameters
$\epsilon_i$; 
(ii) the trilinear model where we introduce five input 
parameters $\lambda'_i$ and $\lambda_i$.

\medskip

$\bullet$ \underline{Bilinear model:  a scheme for the SMA solution.}

\begin{table}
\begin{center}
\begin{tabular}{c|ccc}
\hline
Set1 &  $\tan\beta=10$ &  $\Lambda_S=40$ TeV &   $M_m=150$ TeV  \\
\hline
$\epsilon^0_i$   &
   $7.53\times10^{-6}$ & $2.51\times10^{-4}$ & $2.51\times10^{-4}$ \\
$\tilde{\lambda}'_i$   &
   $1.40\times10^{-6}$ & $4.68\times10^{-5}$ & $4.68\times10^{-5}$ \\
$\tilde{\lambda}_i$   &
   $7.60\times10^{-7}$ & $2.53\times10^{-5}$ & 0     \\
$\xi^0_i$ & 
   $2.40\times10^{-7}$& $8.01\times10^{-6}$ & $7.86\times10^{-6}$ \\
$\xi_i$ & 
   $1.81\times10^{-7}$& $6.06\times10^{-6}$ & $6.63\times10^{-6}$ \\
\hline
BR  &  $e$  & $\mu$   & $\tau$   \\
\hline
$\nu jj$ &   &  $3.58\times10^{-1}$ & \\
$l^\pm_i jj$ & 
  $1.10\times10^{-6}$  & $1.22\times10^{-3}$ & $1.18\times10^{-3}$  \\
$\nu l_i^\pm \tau^\mp$  &
   $4.56\times10^{-4}$  & $3.12\times10^{-1}$ & $3.27\times10^{-1}$  \\
\hline
       & $m_{\chi^0_1}$=49 GeV
     & \hfill $\Gamma=$& $7.18\times10^{-14}$ GeV  \\
\hline
\end{tabular}
$$
\begin{array}{l}
 (\Delta m^2_{31},~ \Delta m^2_{21})=
     (2.5\times10^{-3},~6.1\times10^{-6})~ \mbox{eV}^2  \cr
  (\sin^22\theta_{atm},~ \sin^22\theta_{sol},~ \sin^22\theta_{chooz})
 =(0.99,~0.0018,~0.0017) 
\end{array} 
$$
\caption{A bilinear model with the input parameters, $\tan\beta$, 
$\Lambda_S$, $M_m$ and $\epsilon^0_i$, allowing for the SMA solution.
The values of $\epsilon_i^0$ are set at the mediation scale $M_m$.
The effective trilinear/bilinear R-parity violating parameters,
$\tilde{\lambda}'_i$, $\tilde{\lambda}_i$/$\xi^{(0)}_i$ defined 
in the text, are shown in the upper part. 
Here, $\xi^0_i$ and $\xi_i$ are the tree-level and one-loop 
improved values, respectively.
In the lower part are shown  
the important branching ratios of the LSP with mass 
$m_{\tilde{\chi}^0_1}$ and its total decay rate $\Gamma$.
The three columns correspond to the lepton flavors,
$i=e,\mu$ and $\tau$, respectively.  For the mode $\nu jj$,
we do not distinguish the neutrino flavors.
The resulting neutrino oscillation parameters are presented 
in the last two lines.  
}
\end{center}
\end{table}
      
The bilinear model with the universality condition is known to 
accommodate only the small mixing angle solution (SMA) of solar neutrino
oscillations \cite{CK,valle1}, which is now strongly disfavored by the
recent SNO data \cite{sol-exp}.  This model is an attractive option as
it is the minimal R-parity violating model and provides fairly neat
correlations between the neutrino oscillation parameters and 
the collider signatures.  In this scheme, the effective trilinear
couplings are given by $\tilde{\lambda}'_i=\epsilon_i h_b$ and 
$\tilde{\lambda}_i=\epsilon_i h_\tau$ in Eq.~(\ref{rgesol}) and thus
both the tree and the loop mass matrix takes the form $M^\nu_{ij}
\propto \epsilon_i \epsilon_j$.  The other flavor dependence comes
from the $h_b,h_\tau$ Yukawa coupling effects, which is weak unless
$\tan\beta$ is very large.  Now that the  relation Eq.~(\ref{bicase}) is 
applied here, the determination of the overall size of $\xi$  in 
Eq.~(\ref{xicb}) and two mixing angles $\theta_{23}$ and 
$\theta_{13}$ in Eq.~(\ref{twoangles}) holds almost precisely.
Thus, the atmospheric and CHOOZ neutrino experiments require
$|\xi_1| \ll |\xi_2| \approx |\xi_3|$ which can be  directly 
translated  to the condition $|\epsilon_1| \ll |\epsilon_2| \approx 
|\epsilon_3|$.  This leads to the neutrino mass matrix structure;
$M^{\nu}_{11} < M^\nu_{12} <M^\nu_{22,23,33}$.  As a consequence,
only a small mixing angle for the solar neutrino oscillation
can be accounted for in the bilinear model.  
Indeed, the solar mixing angle $\theta_{12}$ 
is almost fixed by the relation $\tan\theta_{12}
\approx \tan\theta_{13}$, and thus we get the relation 
$\sin^22\theta_{sol} \approx \sin^22\theta_{chooz}$.

Given $\Delta m^2_{32} \approx 3\times10^{-3}$ eV$^2$ for the 
atmospheric neutrino oscillation, Eq.~(\ref{smallratio}) tells us that
$\Delta m^2_{21}< 3\times10^{-9} t_\beta^2$ eV$^2$. 
This estimation is by no means exact but can show some qualitative 
features.  For instance, it  implies that  the right value of 
$\Delta m^2_{21} \sim 5\times10^{-6}$ eV$^2$ is hardly achieved with
small $\tan\beta$ in the GMSB models under consideration.  
In our numerical calculation,  we looked for the SMA solutions varying
the parameters $\tan\beta$, $\Lambda_S$ and $M_m$ as well as two
R-parity violating parameters $\epsilon^0_1$ and $\epsilon^0_2$ 
defined at the scale $M_m$ while keeping $\epsilon^0_2=\epsilon^0_3$.   
We could find a reasonable parameter space only 
for $\tan\beta\approx 10-25$,   
limiting ourselves to $\tan\beta < 25$ because the (right-handed) 
stau becomes the LSP for larger $\tan\beta$. 

\begin{table}
\begin{center}
\begin{tabular}{c|ccc}
\hline
Set2 &  $\tan\beta=25$ &  $\Lambda_S=45$ TeV &   $M_m=90$ TeV  \\
\hline
$\epsilon^0_i$   &
   $8.94\times10^{-7}$ & $2.98\times10^{-5}$ & $2.98\times10^{-5}$ \\
$\tilde{\lambda}'_i$   &
   $4.25\times10^{-7}$ & $1.42\times10^{-5}$ & $1.42\times10^{-5}$ \\
$\tilde{\lambda}_i$   &
   $2.30\times10^{-7}$ & $7.67\times10^{-6}$ & 0     \\
$\xi^0_i$ & 
   $4.36\times10^{-7}$& $1.45\times10^{-5}$ & $1.45\times10^{-5}$ \\
$\xi_i$ & 
   $4.04\times10^{-7}$& $1.35\times10^{-5}$ & $1.28\times10^{-5}$ \\
\hline
BR  &  $e$  & $\mu$   & $\tau$   \\
\hline
$\nu jj$ &   &  $1.11\times10^{-1}$ & \\
$l^\pm_i jj$ & 
  $1.55\times10^{-6}$  & $1.72\times10^{-3}$ & $1.72\times10^{-3}$  \\
$\nu l_i^\pm \tau^\mp$  &
   $6.22\times10^{-4}$  & $4.07\times10^{-1}$ & $4.07\times10^{-1}$  \\
\hline
       & $m_{\tilde{\chi}^0_1}$=58 GeV
     & \hfill $\Gamma=$& $4.30\times10^{-14}$ GeV  \\
\hline
\end{tabular}
$$
\begin{array}{l}
 (\Delta m^2_{31},~ \Delta m^2_{21})=
     (3.0\times10^{-3},~4.1\times10^{-6})~ \mbox{eV}^2  \cr
 (\sin^22\theta_{atm},~ \sin^22\theta_{sol},~ \sin^22\theta_{chooz})
 =(0.99,~0.0019,~0.0017)
\end{array}
$$
\end{center}
\caption{ Same as in Table 1 but with $\tan\beta=25$.}
\end{table}

In Tables 1 and 2, we present two typical sets of parameters accommodating
the SMA solution and the other neutrino data.  In the tables, 
$\epsilon^0_i$ denote input values set at the scale $M_m$ where
supersymmetry breaking is mediated.  
As can be seen, the effective couplings $\tilde{\lambda}'_i$ and 
$\tilde{\lambda}_i$ are much larger than $\xi_i c_\beta$ and the 
dominant decay modes are $\nu jj$ and $\nu ll$ where the diagrams with 
the exchanges of sneutrinos and charged sleptons give main contributions.  
One of its consequence is that the total decay rate is larger than 
$10^{-14}$ GeV  making the decay length smaller than  a few $cm$.
We have checked that the total decay rate is in the region of 
$\Gamma\sim 5\times10^{-14}$ GeV for the LSP mass 
$m_{\tilde{\chi}^0_1}=25-80$ GeV. That is, the decay length is
around $\tau \sim 0.4\, cm$.  This has to be contrasted
with the supergravity case \cite{valle2} where one typically gets
$\tau > 1\,cm$ for a light LSP.
It is also worthwhile to note that the SMA solution requires 
$\xi_1/\xi_2 \approx 0.03 \approx \epsilon_1/\epsilon_2$ 
and thus the bilinear model typically predicts the following relation:
\begin{equation}
 10^3 \mbox{BR}(\nu e^\pm \tau^\mp) \sim  
  \mbox{BR}(\nu \mu^\pm \tau^\mp) \approx  
  \mbox{BR}(\nu \tau^\pm \tau^\mp)  \,.
\end{equation}
As pointed out in Ref.~\cite{valle2}, the modes $l_ijj$ 
are of a great interest.  Their decay rates are dominated by the  
$W$ exchange
diagrams as the contribution of the largest coupling 
$\tilde{\lambda}'_{i33}$ giving $\tilde{\chi}^0_1 \to l_i t \bar{b}$ is 
kinematically forbidden and the coupling $\tilde{\lambda}'_{i22} 
= \epsilon_i h_s$ gives the sub-leading 
effect compared to $\xi_i c_\beta$ which enters the $\chi$-$l$-$W$ vertices.  
Therefore, the ratio
\begin{equation}
 \mbox{BR}(ejj)\;:\;
 \mbox{BR}(\mu jj)\;:\;
 \mbox{BR}(\tau jj)
\end{equation}
is almost same as the ratio $\xi_1^2$ : $\xi_2^2$ : $\xi_3^2$
to determine $\theta_{atm}$ and $\theta_{chooz}$ through Eq.~(23)
as in the case of $m_{\tilde{\chi}^0_1} > M_W$. 
Here, we remark that 
the branching fraction $\mbox{BR}(ejj)$ is 
too small to be observed in the future linear colliders.
Assuming the integrated luminosity 1000 fb$^{-1}$ per year,
the branching ratios below $10^{-5}$ would not be feasible \cite{valle2}.  
However,  the measurement of BR($ejj$) $\ll$ 
$\mbox{BR}(\mu jj) \approx \mbox{BR}(\tau jj)$  will provide a robust
test for the bilinear model.

\medskip

$\bullet$ \underline{Trilinear model: a scheme for the LMA solution.}

Let us now consider a more general situation that both bilinear 
and trilinear R-parity violating terms are present.  In this case,
it is convenient to rotate away the supersymmetric bilinear terms 
$\epsilon_i$ to the trilinear couplings as we defined the
effective ones in the previous sections.  In this way, we are allowed to
introduce only five couplings, $\tilde{\lambda}'_i$ and $\tilde{\lambda}_i$ 
which are related to the third generation quarks and leptons.
This would be the simplest trilinear model.

The trilinear model provides a possibility to realize 
the LMA solution which is most favored at present.
As discussed before,  in order to get the LMA solution,
sizable contributions to $M^\nu_{11,12,22}$ are needed 
to enlarge the solar neutrino mixing while keeping the hierarchy of
$M_{ij}< M_{i3,33}$ to realize the large atmospheric neutrino mixing.  
From the numerical calculation scanning the five trilinear parameters,
we find that the LMA solution is realized  if
$\tilde{\lambda}_{1,2} \sim 5 \tilde{\lambda}'_{2,3}$.
The conditions, $\tilde{\lambda}_{1}\sim \tilde{\lambda}_{2}$ and 
$\tilde{\lambda}'_{2}\sim \tilde{\lambda}'_{3}$, are needed to 
get two large mixing angles,  while the small CHOOZ angle requires
$\tilde{\lambda}'_{1} < 0.2 \tilde{\lambda}'_{2,3}$.
Under such conditions, we could not find any restrictions on the GMSB
input parameters $\tan\beta$, $\Lambda_S$ and $M_m$.

\begin{table}
\begin{center}
\begin{tabular}{c|ccc}
\hline
Set3 &  $\tan\beta=5$ &  $\Lambda_S=40$ TeV &   $M_m=80$ TeV  \\
\hline
$\tilde{\lambda}'_i$   &
   $1.07\times10^{-7}$ & $1.07\times10^{-5}$ & $0.96\times10^{-5}$ \\
$\tilde{\lambda}_i$   &
   $4.07\times10^{-5}$ & $4.07\times10^{-5}$ & 0     \\
$\xi^0_i$ & 
   $2.66\times10^{-7}$& $2.99\times10^{-6}$ & $2.48\times10^{-6}$ \\
$\xi_i$ & 
   $2.57\times10^{-7}$& $2.67\times10^{-6}$ & $2.66\times10^{-6}$ \\
\hline
BR  &  $e$  & $\mu$   & $\tau$   \\
\hline
$\nu jj$ &   &  $7.92\times10^{-3}$ & \\
$l^\pm_i jj$ & 
  $5.17\times10^{-7}$  & $6.55\times10^{-5}$ & $4.49\times10^{-5}$  \\
$\nu l^\pm_i \tau^\mp$  &
   $2.35\times10^{-1}$  & $2.35\times10^{-1}$ & $5.22\times10^{-1}$  \\
\hline
       & $m_{\tilde{\chi}^0_1}$=46 GeV
     & \hfill $\Gamma=$& $1.20\times10^{-13}$ GeV  \\
\hline
\end{tabular}
\end{center}
$$ \begin{array}{l}
 (\Delta m^2_{31},~ \Delta m^2_{21})=
     (3.1\times10^{-3},~5.0\times10^{-5})~ \mbox{eV}^2  \cr
 (\sin^22\theta_{atm},~ \sin^22\theta_{sol},~ \sin^22\theta_{chooz})
 =(0.99,~0.80,~0.0001)
\end{array} $$
\caption{A trilinear model realizing the LMA solution. Here
the couplings $\tilde{\lambda}'_i$ and $\tilde{\lambda}_i$ can be 
considered as input parameters defined at the weak scale.  
The rests are the same as in the previous tables.}
\end{table}
      
In Tables 3 and 4, we present two examples allowing the LMA solution
for $\tan\beta = 5$ and 25, respectively. 
We fixed $M_m = 2 \Lambda_S$.
The total decay rate is found to be in the vicinity of $\Gamma \sim 
5\times10^{-13}$ GeV for all the LSP mass below 80 GeV.   
Therefore, the decay length is of the order $0.4\; mm$.
This enhancement compared to the bilinear case
is due to the largeness of the $\tilde{\lambda}_i$ couplings which also 
make the modes $\nu ll$ dominating.  
As a consequence, we infer the following distinct feature 
of the LMA solution;
\begin{equation}
\mbox{BR}(\nu e \bar{\tau}) \sim \mbox{BR}(\nu \mu \bar{\tau}) \sim 
\mbox{BR}(\nu \tau \bar{\tau})
\end{equation}
with the individual branching ratio is larger than 10\%.
The relation for the atmospheric neutrino mixing angle in 
Eq.~(\ref{twoangles}) is not  as exact as in the bilinear case,
but it  still  holds to a good approximation as can be seen from the 
tables.   On the other hand,  the expression for the CHOOZ angle 
determined from $\xi_i$'s is not applicable any more and 
we cannot draw any conclusive prediction for the value of it.
Still, the relation, BR($ejj$) $\ll$ BR($\mu jj$) $\sim$ BR($\tau jj$),
holds to a certain degree, but these branching fractions become  
as small as $10^{-5}$ making it difficult to be measured
in the planned colliders.

\begin{table}
\begin{center}
\begin{tabular}{c|ccc}
\hline
Set4 &  $\tan\beta=25$ &  $\Lambda_S=40$ TeV &   $M_m=80$ TeV  \\
\hline
$\lambda'_i$   &
   $7.45\times10^{-8}$ & $4.48\times10^{-6}$ & $7.43\times10^{-6}$ \\
$\lambda_i$   &
   $1.61\times10^{-5}$ & $2.82\times10^{-5}$ & 0     \\
$\xi^0_i$ & 
   $4.32\times10^{-7}$& $7.09\times10^{-6}$ & $1.09\times10^{-5}$ \\
$\xi_i$ & 
   $1.62\times10^{-6}$& $1.02\times10^{-5}$ & $1.28\times10^{-5}$ \\
\hline
BR  &  $e$  & $\mu$   & $\tau$   \\
\hline
$\nu jj$ &   &  $1.54\times10^{-3}$ & \\
$l^\pm_i jj$ & 
  $7.08\times10^{-8}$  & $1.91\times10^{-5}$ & $4.53\times10^{-5}$  \\
$\nu l_i \bar{l}_3$  &
   $1.15\times10^{-1}$  & $3.53\times10^{-1}$ & $5.31\times10^{-1}$  \\
\hline
       & $m_{\tilde{\chi}^0_1}$=50 GeV
     & \hfill $\Gamma=$& $5.62\times10^{-13}$ GeV  \\
\hline
\end{tabular}
\end{center}
$$ \begin{array}{l}
 (\Delta m^2_{31},~ \Delta m^2_{21})=
     (3.0\times10^{-3},~5.0\times10^{-5})~ \mbox{eV}^2  \cr
 (\sin^22\theta_{atm},~ \sin^22\theta_{sol},~ \sin^22\theta_{chooz})
 =(0.91,~0.84,~0.16)
\end{array} $$
\caption{Same as in Table 3 with $\tan\beta=25$.}
\end{table}

\section{Conclusion}

The supersymmetric standard model without R-parity is an attractive
framework for the neutrino masses and mixing as certain neutrino
oscillation parameters can be probed by measuring the decay length 
and various branching fractions of the neutralino LSP in the 
future colliders experiments.  Taking two simple models of R-parity 
violation, the bilinear model with three input parameters 
and the trilinear model with five parameters,  we analyzed the
neutrino mass matrix which explains both the atmospheric and solar
neutrino data and its consequences on collider searches.   
One of our basic assumptions is the universality of soft terms 
for which we considered gauge mediation models of supersymmetry breaking.  
A notable consequence of such an assumption is that the LSP (lighter than
the $W$ boson) decays mainly through the (effective) trilinear 
couplings $\lambda'_i$ and $\lambda_i$ and its decay length is 
found to be in the ballpark of $ \tau \sim 0.1\,cm$.

The observation of the  decay modes $\nu l^\pm_i l^\mp_j$ and $l_i^\pm jj$ 
will be important as they reflect the lepton number violating structure
of a certain model.
The bilinear model which can accommodate only the SMA solution of the
solar neutrino oscillation predicts the relation,
$10^3$ BR($\nu e^\pm \tau^\mp$) $\sim$
BR($\nu \mu^\pm \tau^\mp$) $\approx$ BR($\nu \tau^\pm \tau^\mp$).
The dominant decay modes are found to be $\tilde{\chi}^0_1 \to$ 
$\nu \mu^\pm \tau^\mp$, $\nu \tau^\pm \tau^\mp$ and $\nu jj$, which
are all of the order 10\%.
The trilinear model can realize the 
strongly-favored LMA solution which can be tested by observing 
the dominant decay modes, $\tilde{\chi}^0_1 \to$  
$\nu e^\pm \tau^\mp$, $\nu \mu^\pm \tau^\mp$ and 
$\nu \tau^\pm \tau^\mp$, satisfying the relation, 
BR($\nu e^\pm \tau^\mp$) $\sim$ BR($\nu \mu^\pm \tau^\mp$) $\sim$ 
BR($\nu \tau^\pm \tau^\mp$).
In both cases,  the relation,
BR($ejj$) $\ll$ BR($\mu jj$) $\sim$ BR($\tau jj$), should hold
to be consistent with the atmospheric and CHOOZ neutrino data.

\medskip

{\bf Acknowledgment}:  EJC and JDP are supported by the KRF grant
No. 2001-003-D00037.

\end{document}